\documentclass[sigconf]{acmart}

\AtBeginDocument{%
  \providecommand\BibTeX{{%
    \normalfont B\kern-0.5em{\scshape i\kern-0.25em b}\kern-0.8em\TeX}}}

\copyrightyear{2024}
\acmYear{2024}
\setcopyright{acmlicensed}\acmConference[CIKM '24]{Proceedings of the 33rd ACM International Conference on Information and Knowledge Management}{October 21--25, 2024}{Boise, ID, USA}
\acmBooktitle{Proceedings of the 33rd ACM International Conference on Information and Knowledge Management (CIKM '24), October 21--25, 2024, Boise, ID, USA}
\acmDOI{10.1145/3627673.3679831}
\acmISBN{979-8-4007-0436-9/24/10}



\NewDocumentCommand\githubicon{}{\includegraphics[scale=0.025]{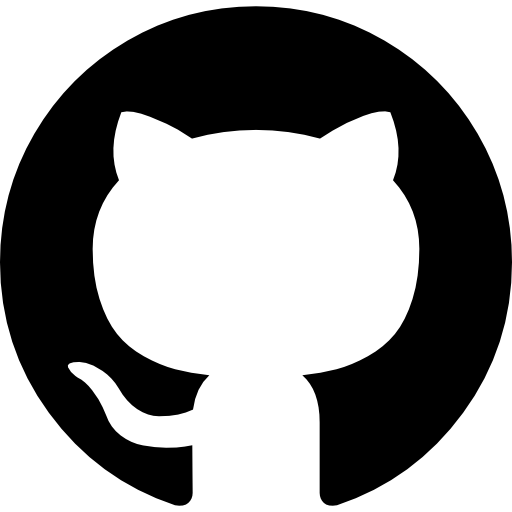}}
\usepackage{balance}
\usepackage{makecell}
\usepackage{graphicx}
\usepackage{subcaption}
\usepackage{tabularx}
\usepackage{multirow}
\usepackage{arydshln}
\usepackage{hyperref}
\usepackage{hyperxmp}
\usepackage{enumitem}
\usepackage[nameinlink]{cleveref}
\crefformat{section}{\S#2#1#3} 
\crefformat{subsection}{\S#2#1#3}
\crefformat{subsubsection}{\S#2#1#3}
\crefformat{paragraph}{\S#2#1#3}
\crefname{figure}{Figure}{Figures}
\crefname{table}{Table}{Tables}
\begin{document}

\title{Navigating the Landscape of Reproducible Research: A Predictive Modeling Approach}

\author{Akhil Pandey Akella}
\affiliation{%
  \institution{Northern Illinois University}
  \city{Dekalb}
  \state{Illinois}
  \country{USA}}
\affiliation{%
  \institution{Northwestern University}
  \city{Evanston}
  \state{Illinois}
  \country{USA}}
\email{aakella@niu.edu}

\author{Sagnik Ray Choudhury}
\affiliation{%
  \institution{University of North Texas}
  \city{Denton}
  \state{Texas}
  \country{USA}}
\email{sagnikrayc@gmail.com}

\author{David Koop}
\affiliation{%
  \institution{Northern Illinois University}
  \city{Dekalb}
  \state{Illinois}
  \country{USA}}
\email{dakoop@niu.edu}

\author{Hamed Alhoori}
\affiliation{%
  \institution{Northern Illinois University}
  \city{Dekalb}
  \state{Illinois}
  \country{USA}}
\email{alhoori@niu.edu}

\renewcommand{\shortauthors}{Akhil Pandey Akella, Sagnik Ray Choudhury, David Koop, \& Hamed Alhoori}

\begin{abstract}
The reproducibility of scientific articles is central to the advancement of science. Despite this importance, evaluating reproducibility remains challenging due to the scarcity of ground truth data. Predictive models can address this limitation by streamlining the tedious evaluation process. Typically, a paper's reproducibility is inferred based on the availability of artifacts such as code, data, or supplemental information, often without extensive empirical investigation. To address these issues, we utilized artifacts of papers as fundamental units to develop a novel, dual-spectrum framework that focuses on author-centric and external-agent perspectives. We used the author-centric spectrum, followed by the external-agent spectrum, to guide a structured, model-based approach to quantify and assess reproducibility. We explored the interdependencies between different factors influencing reproducibility and found that linguistic features such as readability and lexical diversity are strongly correlated with papers achieving the highest statuses on both spectrums. Our work provides a model-driven pathway for evaluating the reproducibility of scientific research. The code, methods, and artifacts for our study are publicly available. \footnote{\githubicon\  \url{https://github.com/reproducibilityproject/NLRR/}}
\end{abstract}

\maketitle

\section{Introduction}

The abundance of open-source libraries, version control frameworks, and publicly-available, archived datasets has made it easier than ever to ensure transparency in the scientific process. However, this increased attention on research reproducibility \cite{ivie2018reproducibility,national2019reproducibility} has not necessarily driven the scholarly community to implement more transparent measures to make their work fully reproducible. Instead, an inverse phenomenon is observed: surveys indicate that scientists often believe many scholarly articles are irreproducible \cite{baker2016reproducibility}, a sentiment that spans multiple fields \cite{fanelli2018science}.

Given the existing perception, it is crucial to develop a data-driven approach that can establish trust in the reproducibility of scientific papers. The reproducibility of research papers is a complex issue \cite{bajpai2017challenges,cockburn2020threats}. For example, consider a computational paper that researchers fail to reproduce despite the publicly available code and data, possibly due to the unavailability of specific libraries used in the original code. Such a paper should not be categorized alongside those that made no effort to ensure reproducibility. Therefore, reproducibility should be viewed as a \textit{spectrum} rather than a binary classification. By acknowledging varying degrees of reproducibility, we can elevate trust across the board and help identify common factors that contribute to reproducible research. This refined approach reduces the collective burden on conferences, journals, publishers, and the research community at large.

The initial step in constructing a reproducibility spectrum is collecting existing ground truth about signals that indicate reproducible work. This can include meta-studies confirming the reproducibility of existing research \cite{sweeney2017methods}, citations where methodologies have been re-implemented \cite{obadage2024can}, and reproducibility challenges hosted by premier conferences \cite{arguello2016report,lin2016toward,clancy2019sigir,pineau2021improving}. While these serve as proxy measures for reproducibility, establishing definitive ground truth for the reproducibility of scholarly work is challenging and limited to a few sources. For example, conferences such as \textit{OOPSLA, PLDI, and ISSTA} have conducted reproducibility reviews \cite{boisvert2016incentivizing} to formally evaluate software artifacts and data. The practice of evaluating artifacts was first established at \textit{SIGMOD} 2008 \cite{pawlik2019link,chirigati2016reprozip}, and various sub-disciplines within the Association for Computing Machinery (\textit{ACM}) have since adopted similar policies to audit artifacts. Collecting signals from these efforts was fundamental for establishing the \textit{ACM} badging process. In this process, a paper may receive badges such as \textit{Artifacts Available}, \textit{Artifacts Evaluated-Reusable}, and \textit{Results Reproduced}. This policy acknowledges the researchers' efforts and incentivizes reproducibility.

While efforts like ACM Badging encourage the creation of reproducible research, the current system places a significant burden on the committees that evaluate artifact availability and reproducibility. However, the specific procedures for awarding reproducibility badges can vary across venues. Moreover, much of the literature on estimating and understanding reproducibility has relied on traditional modeling \cite{yang2020estimating} and cohort-based statistical analysis \cite{raff2019step}. While valuable, these approaches cannot scale effectively -- assistance of automated systems such as predictive models is needed.

In this paper, we present a predictive modeling study utilizing a novel joint spectrum on reproducibility. This spectrum consists of an author-centric framework ($A$) and an external-agent framework ($E$). The author-centric framework identifies efforts made by authors to enhance the transparency and accessibility of their papers and is composed of three categories. The external-agent framework characterizes the success of external reviewers' efforts to reproduce a paper and is composed of four categories.

In summary, our contributions are: First, we present a novel approach to characterize reproducible research. Second, we analyze various features extracted from the text and metadata of papers to understand their relevance to reproducibility. Finally, we build an interpretable model for predicting how reproducible a paper might be. Unlike the current ad-hoc method of assigning subjective scores by reviewers, our approach is more systematic and data-driven.

We acknowledge the ethical and moral implications of utilizing a predictive model to assist in evaluating the quality and reproducibility of research papers. However, our goal in this study is to provide empirical evidence to support the use of such models and to identify crucial aspects influencing a paper's reproducibility assessment. We envision that the results of these models will complement and support reviewers in navigating the landscape of reproducible research rather than replacing human judgment.

\section{Background and Related work}
\label{sec:background}

Researchers from the University of Arizona \cite{collberg2015repeatability, collberg2016CACM} analyzed data on computer systems research in an attempt to measure and understand reproducibility. Although these efforts didn't generate a conclusive hypothesis, they were instrumental in initiating a process to observe the willingness of computer science researchers to share code and data. Examining the conflicting attitudes of researchers towards reproducibility \cite{baker2016reproducibility} provided insights into the frequency of successful and unsuccessful replications at both individual and disciplinary levels. The scholarly community acknowledged the reproducibility crisis, and there has been momentum for initiatives such as creating a manifesto on reproducibility \cite{munafo2017manifesto} and estimating reproducibility rates \cite{open2015estimating}.

Reproducibility has been formalized and recognized by various players involved in the scholarly publication process such as publishers, conferences, and peer reviewers. This recognition led to the establishment of funding programs such as DARPA's SCORE (Systematizing Confidence in Open Research and Evidence), which encourages researchers to develop assessment strategies to measure replication and reproduction efforts that are central to the scientific process. Additionally, many organizations introduced reproducibility checklists, most prominently ACM's rollout of Artifact Review and Badging \footnote{\url{https://www.acm.org/publications/policies/artifact-review-badging}} to address reproducibility and enhance research integrity across computational disciplines. 

Literature that aligns with our goals for measuring and estimating reproducibility includes terminology papers \cite{gundersen2018state, gundersen2021fundamental, fanelli2018science, raff2023siren}, statistical studies quantifying factors influencing reproducibility \cite{raff2019step, yildiz2021reproducedpapers}, and predictive modeling studies \cite{yang2020estimating, raff2023does, salsabil2022study, rajtmajer2022synthetic}. While these studies set an appropriate foundation, they fall short in one or more aspects to be considered conclusive in identifying reproducible works preemptively. These limitations include:

\begin{enumerate}[leftmargin=*,itemsep=0pt]
    \item \textbf{Lack of comprehensive methodology}: Most quantitative studies on reproducibility approach the analysis from a single perspective, often relying on correlations, statistical tests, predictive models, or user surveys. Identifying the reproducibility of a paper requires a comprehensive methodology capable of detecting a wide range of signals.
    \item \textbf{Potential impact on unseen data}: Understanding the reproducibility of scholarly works requires high standards of data curation. Given the limited number of works verified as reproducible, generalization becomes a challenge. It is crucial to outline the broader impact and limitations of the quantitative analysis on unseen data to validate the findings effectively. 
    \item \textbf{Optimal balance on subjective vs. objective attributes}: Factors such as field of study, discipline, and venue significantly influence the structure of scientific research and the methodologies used in experiments. It is essential to strike an optimal balance between subjective and objective features when analyzing the causes of reproducible outcomes to ensure that findings about reproducibility are generalizable.  
\end{enumerate}

Given the significant challenges in gathering data on reproducibility, especially in computational science, our current study can serve as a primer for discussions on this topic. Building on related works \cite{raff2019step, yang2020estimating, akella2022reproducibility}, our study provides a comprehensive modeling approach to identify crucial aspects of papers that can predict whether it would be reproducible.

\section{Building the Dataset}

Our goal is to create a dataset that can be quantitatively analyzed in relation to artifacts and reproducibility. To achieve this, we collected papers from the ACM Digital Library because it is a singular comprehensive source with detailed information about the artifacts and reproducibility of scholarly articles. The ACM introduced the Artifact Reviewing and Badging policy, which assigns badges to indicate when publications have been successfully reproduced. These badges include:

\begin{enumerate}[leftmargin=*,itemsep=0pt]
    \item \textbf{Artifacts Available}: Assigned when papers include artifacts that have been made permanently retrievable.
    \item \textbf{Artifacts Evaluated and Functional} or \textbf{Artifacts Evaluated and Reusable}: Assigned when the artifacts have been reviewed and audited.
    \item \textbf{Results Reproduced}: Given when the primary findings of the publication have been validated and independently verified in a later investigation by a person or group other than the authors \textit{without the use of author-supplied artifacts}.
\end{enumerate}

\subsection{Data Collection}

Our data collection process involved the following steps:

\begin{enumerate}[leftmargin=*,itemsep=0pt]
    \item Using the ACM digital library advanced search endpoint \footnote{\url{https://dl.acm.org/search/advanced}} to list all scholarly articles in the ACM full-text collection that have received the \textit{Results Reproduced} badge.   
    \item Conducting separate searches using the same ACM digital library advanced search endpoint for articles with each of the following badges: \textit{Artifacts Available}, \textit{Artifacts Evaluated and Functional}, and \textit{Artifacts Evaluated and Reusable}.
    \item Identifying the venues of articles with the ``Results Reproduced'' badge, and collecting unbadged articles from the same venues that were published in the same respective issue/year.
\end{enumerate} 

This resulted in an initial collection of just over three thousand badged articles. To maintain relevance, we included only papers published between 2016 and 2023, aligning with the timeframe when the ACM Badging policy was implemented. By filtering the samples based on full-text availability and publication date, we finalized a dataset of 2,659 articles. These articles were categorized as either \textit{Artifacts Available}, \textit{Artifacts Evaluated \& Functional}, \textit{Artifacts Evaluated \& Reusable}, \textit{Results Reproduced}, or \textit{Unbadged}. \textit{Unbadged} refers to papers from the same venues and years as \textit{Results Reproduced} papers that were manually collected and included because the authors chose not to submit them for artifact \& reproducibility evaluation.

\begin{figure}[htbp]
  \centering
    \includegraphics[width=\linewidth]{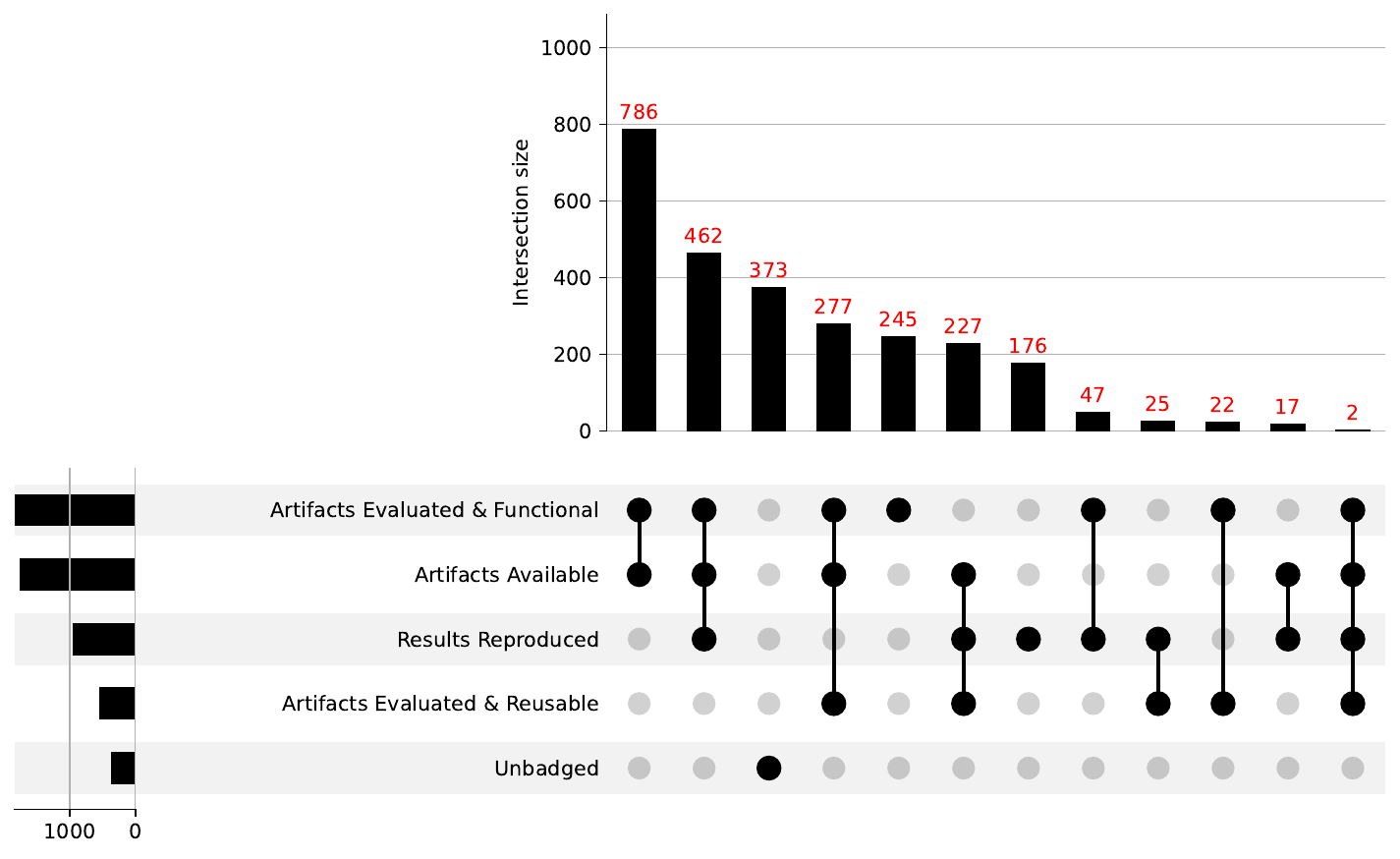}
    \caption{Visualization of badge category overlaps for the scholarly articles in our dataset.}
    \label{fig:acm_badges_data}
    \Description{Visualizing badge category overlaps for the ACM articles}
\end{figure}

The distribution of badges and the overlap between categories are illustrated in Fig. \ref{fig:acm_badges_data}. Interestingly, many badged articles have multiple badge combinations.
Fig.~\ref{fig:acm_badges_data} shows that articles with the badges \textit{Artifacts Available}, and \textit{Artifacts Evaluated \& Functional} have the largest intersection with 786 articles. In contrast, only 2 articles have all the badges. Furthermore, most reproducible articles tend to overlap with the \textit{Artifacts Available} and \textit{Artifacts Evaluated \& Functional} categories. Noticeably, the Unbadged set has the highest unique category count, with 373 articles.

\begin{figure}[htb]
    \includegraphics[width=\linewidth]{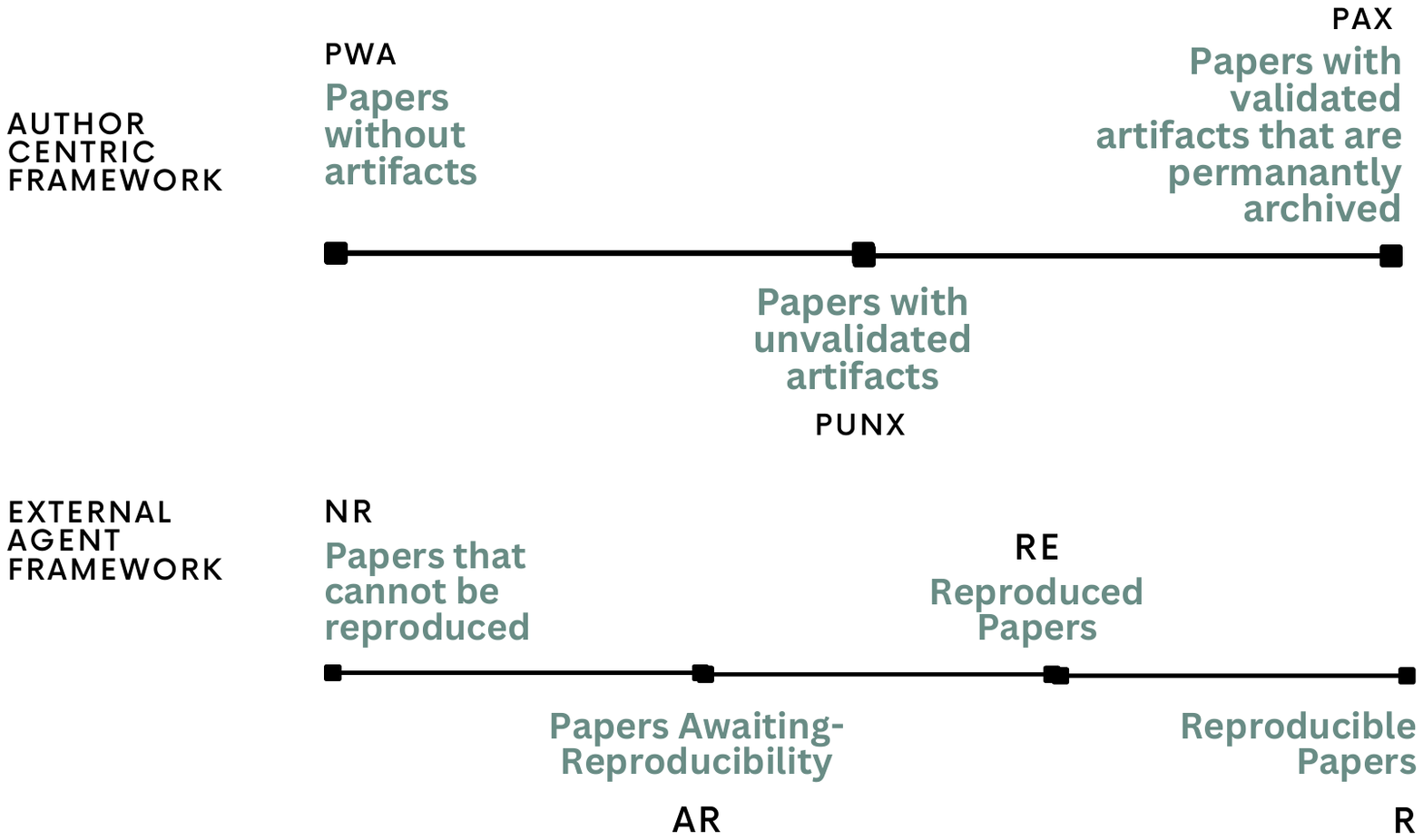}
    \caption{Joint framework to assess reproducibility levels in scientific papers.}
    \label{fig:reproducibility_joint_spectrum}
\end{figure}

\section{Reproducibility Spectrum}
\label{sec:spectrum}

We introduce a joint spectrum for evaluating reproducibility in scientific papers as illustrated in Fig.~\ref{fig:reproducibility_joint_spectrum}. This spectrum is a result of a data-driven, iterative development process. Initially, our concept of the reproducibility spectrum categorized works as reproducible or non-reproducible. However, this simplistic approach failed to capture the nuances of scientific papers, as highlighted in Fig.~\ref{fig:acm_badges_data}. Our data collection showed a much more complex landscape with interesting sub-categories of papers. This process revealed the importance of artifacts as a critical unit for assessing reproducibility. We finally constructed a version of the spectrum that is composed of an author-centric framework and an external agent framework. The author-centric framework focuses on the quality and availability of artifacts provided by the authors. It recognizes the varying degrees of effort authors put into making their work reproducible. The external-agent framework captures the external validation of a paper's reproducibility based on the available artifacts. By separating these aspects, we were able to represent the multifaceted nature of reproducibility in scientific publications.

\subsection{Author-Centric Framework}

The author-centric framework broadly captures the varying degrees of effort and commitment authors invest to facilitate reproducibility. The labels within this framework includes $\mathbf{A_i}$:
 \begin{itemize}[leftmargin=*,itemsep=0pt]
    \item $\mathbf{A_{PWA}}$: Papers without artifacts
    \item $\mathbf{A_{PUNX}}$: Papers with unvalidated artifacts
    \item $\mathbf{A_{PAX}}$ Papers with validated artifacts that are permanently archived
\end{itemize} 

The key difference between $\mathbf{A_{PUNX}}$ and $\mathbf{A_{PAX}}$ is that $\mathbf{A_{PAX}}$ includes validation of the archived artifacts. While $\mathbf{A_{PUNX}}$ may include artifacts that are either archival or non-archival, the crucial difference between it and $\mathbf{A_{PAX}}$ is that they are not validated. Therefore, papers with validated artifacts that are permanently archived are considered the highest standard on our spectrum since they represent papers where authors took the most proactive measures to facilitate reproducibility evaluations.

We used the ACM badge (or the absence of one) to assign labels in the author-centric framework as follows:
\begin{itemize}[leftmargin=*,itemsep=0pt]
     \item $A_{PWA}$: For all unbadged papers.
     \item $A_{PUNX}$: For papers with either the \textit{Artifacts Available} or \textit{Artifacts Evaluated \& Functional} badge.
     \item $A_{PAX}$: For papers with the \textit{Artifacts Evaluated \& Reusable} badge, indicating the highest effort towards permanently archiving the paper's artifacts.
 \end{itemize}

\subsection{External Agent Framework}

The external-agent framework presents the reproducibility evaluation status of a paper based on the information available for an independent team to assess and validate the original study's findings. This framework categorizes papers into the following units $E_i$ on the spectrum: 

\begin{itemize}[leftmargin=*,itemsep=0pt]
    \item $\mathbf{E_{NR}}$: Papers that cannot be reproduced
    \item $\mathbf{E_{AR}}$: Papers awaiting-reproducibility
    \item $\mathbf{E_{Re}}$: Reproduced papers
    \item $\mathbf{E_{R}}$: Reproducible papers
\end{itemize}

There are several points to notice. First, $E_{NR}$ papers lack any artifacts or supplemental information necessary for initiating reproducibility evaluation. Second, papers that are classified as Reproduced $E_{Re}$ or Reproducible $E_{R}$ have obtained their status through voluntary submission of artifacts to an evaluation committee. There is an important distinction between papers labeled $E_{Re}$ and those labeled $E_{R}$, which is based on the archival nature of the artifacts and the reproducibility status. If a paper has $A_{PAX} \cap E_{Re}$, then it is considered $E_{R}$. In contrast, if a paper has been reproduced by any independent team, the assumption of its reproducibility status captured by $E_{Re}$ is based on trust in the independent team's evaluation. In the external-agent framework, we assign labels as follows: 

\begin{itemize}[leftmargin=*,itemsep=0pt]
     \item $E_{NR}$: For all unbadged papers that cannot be reproduced due to a lack of available artifacts.
     \item $E_{AR}$: For papers that have artifacts but that have not yet been reproduced.
     \item $E_{Re}$: For \textit{Results Reproduced} papers that do not have permanently archived artifacts.
     \item $E_{R}$: For papers that have both the \textit{Results Reproduced} badge and the \textit{Artifacts Evaluated \& Reusable} badge.
 \end{itemize}

Moving toward the rightmost end of either spectrum reflects a higher level of effort by the authors. At the same time, the ACM badges have interesting intersections as shown in Fig.~\ref{fig:acm_badges_data}. Specifically, a paper with a ``Results Reproduced'' badge need not have the artifacts available.

\begin{table}
    \centering
    \small
    \caption{Features with their respective categories.}
    \label{tab:feature_details}
    \begin{tabular}{rl}
        \hline
        \textbf{Feature} & \textbf{Category} \\
        \hline
        Number of Algorithms ($X_1$) & Structural \\
        Number of Equations ($X_2$) & Structural \\
        Google Scholar citations ($X_3$) & Scholarly \\
        Availability of reproducibility checklist ($X_4$)& Venue \\
        Mandatory artifact submission for papers ($X_5$) & Venue \\
        Reproducibility Awards ($X_6$) & Venue \\
        Author Correspondence for Reproducibility ($X_7$) & Venue \\
        Mention of Zenodo Artifacts ($X_8$) & Artifact \\
        Mention of GitHub Code Repository ($X_9$) & Artifact \\
        Mention on Papers With Code GitHub Repository ($X_{10}$) & Artifact \\
        Mention on Papers With Code Datasets ($X_{11}$) & Artifact \\
        Mention on Papers With Code Methods ($X_{12}$) & Artifact \\
        Median Readability ($X_{13}$) & Linguistic \\
        Measure of lexical textual diversity ($X_{14}$) & Linguistic \\
        Availability of Funding source ($X_{15}$) & Miscellaneous \\
        Availability of Supplemental information ($X_{16}$) & Miscellaneous \\
        \hline
    \end{tabular}
\end{table}

\section{Pre-processing and Observations}

Previous work \cite{raff2019step} suggests including a wide range of both subjective and objective features to predict reproducibility, whereas the deep learning model from \cite{yang2020estimating} focuses exclusively on the representational power of full-text embeddings. We selected a combination of Structural, Scholarly, Venue, Artifact, Linguistic, and Miscellaneous features, as detailed in \autoref{tab:feature_details}. The Structural, Scholarly, and Linguistic features are numerical, whereas Venue and Miscellaneous features are categorical.

The metadata for each paper was collected from the ACM Digital Library website using a customized web scraper written in Python using the packages Selenium\footnote{\url{https://pypi.org/project/selenium/}}, and BeautifulSoup\footnote{\url{https://pypi.org/project/beautifulsoup4/}}. Additionally, we gathered complete metadata for all articles in our dataset using Allen AI's Academic Graph API (1.0)\footnote{\url{https://api.semanticscholar.org/api-docs/graph}}. We utilized a similar web scraper to gather citations for each paper from Google Scholar, covering citations up to the end of 2023. To gather Miscellaneous and Venue features, we examined the individual article webpages. Miscellaneous features include details about funding and additional supplemental information such as videos, slides, and screen recordings. The Structural and Linguistic features were derived using the full texts of the article, which were processed by passing the PDFs through Allen AI's Science Parse\footnote{\url{https://pypi.org/project/science-parse-api/}}. 

``Readability'' is a linguistic concept that measures how easily a reader can understand a written text. It considers the complexity of vocabulary, sentence structures, and overall text composition. The Median Readability was calculated in two steps. First, we used Python's Textstat \footnote{\url{https://pypi.org/project/textstat/}} package to compute various readability metrics, including the Flesch Reading Ease Score, SMOG Index, Coleman-Liau Index, Automated Readability Index, Dale-Chall Readability Score, Linsear Write Formula, and Fog Scale (Gunning FOG Formula). Then, we calculated a weighted normalized score (ranging from 0 to 1) using the hypothetical minimum and maximum values for all these measures and took the median. Lexical diversity, which reflects the variety and richness of the vocabulary used in a text, was quantified using the Measure of Textual Lexical Diversity (MTLD) \cite{mccarthy2010mtld}. As with readability, we employed the Textstat package to calculate this measure across the full text of each document.

\begin{table}[H]
    \centering
    \small
    \begin{tabular}{lcc}
        \hline
        \textbf{Feature} & \textbf{Statistic} & \textbf{p-value} \\
        \hline
        Median Readability & 0.952613 & 1.565888e-28 \\
        Number of Algorithms & 0.554016 & 1.731140e-63 \\
        Number of Equations & 0.244451 & 8.599141e-74 \\
        Google Scholar citations & 0.100468 & 2.239210e-77 \\
        Measure of lexical textual diversity & 0.855591 & 4.103254e-44 \\
        \hline
    \end{tabular}
    \caption{Shapiro-Wilk Test for assessing the normality of numerical features in scholarly papers.}
    \label{tab:shapiro_test_all_features}
\end{table}

We observed several interesting patterns building our dataset. First, 1.76\% of articles have a dataset mentioned on PapersWithCode, and 1.01\% reference a method on PapersWithCode. Additionally, 9.43\% of articles have an official GitHub repository linked to the paper on PapersWithCode. This information was gathered by cross-referencing Arxiv IDs and paper titles with the PapersWithCode API \footnote{\url{https://paperswithcode.com/api/v1/docs/}}. Further textual analysis revealed that 41.03\% of the articles mention a GitHub repository in the full text (excluding the ``References'' section). We also found that 16\% of the articles reference Zenodo in the full text, pointing to artifacts related to the study. Moreover, 32.49\% of the articles provide supplemental information such as code, audio, or video files on the ACM Digital Library. Finally, 50.1\% of the articles mention funding sources, with the National Science Foundation, Engineering and Physical Sciences Research Council, and Deutsche Forschungsgemeinschaft being the most frequently cited agencies.

\begin{table}[H]
    \centering
    \small
    \begin{tabular}{lcc}
        \hline
        \textbf{Feature} & \textbf{Statistic} & \textbf{p-value} \\
        \hline
        Median Readability & 4.990988  & 6.862850e-03 \\
        Number of Algorithms & 36.773371 & 1.764600e-16 \\
        Number of Equations & 5.258889 & 5.255375e-03 \\
        Google Scholar citations & 1.714010 & 1.803412e-01 \\
        Measure of lexical textual diversity & 1.290552 & 2.752913e-01 \\
        \hline
    \end{tabular}
    \caption{Levene's Test for Homogeneity of Variances grouped by the author-centric framework.}
    \label{tab:levenes_test_auth_framework}
\end{table}

\section{Statistical tests}

The foundation of our predictive modeling study is based on a statistical analysis of the numerical features $\mathbf{X}$ outlined in \autoref{tab:feature_details}. This analysis involved conducting tests for normalization and variance of groups using the Shapiro-Wilk test and Levene's test, followed by a significance test using the Kruskal-Wallis test. Together, these tests ensure the statistical robustness of our feature set by verifying the assumptions of normality and homogeneity of variance, which are important for selecting appropriate predictive models. Additionally, these tests assisted us in discerning significant differences in features observed in both frameworks across different groups of scholarly papers. Finally, these tests guided our choices to pick predictive models that are well-suited to the data distribution.

\begin{table}[H]
    \centering
    \small
    \begin{tabular}{lcc}
        \hline
        \textbf{Feature} & \textbf{Statistic} & \textbf{p-value} \\
        \hline
        Median Readability & 4.153057  & 6.039707e-03 \\
        Number of Algorithms & 29.537040 & 8.830013e-19 \\
        Number of Equations & 6.959335  & 1.158253e-04 \\
        Google Scholar citations & 4.195924  & 5.690491e-03 \\
        Measure of lexical textual diversity & 0.283903  & 8.370575e-01 \\
        \hline
    \end{tabular}
    \caption{Levene's Test for Homogeneity of Variances grouped by the external-agent framework.}
    \label{tab:levenes_test_ext_agent_framework}
\end{table}

The results of the Shapiro-Wilk test for assessing the normality of distributions and Levene's test for evaluating variance across groups are presented in \cref{tab:shapiro_test_all_features,tab:levenes_test_auth_framework,tab:levenes_test_ext_agent_framework}. The Shapiro-Wilk test results indicate that the $p$ values from \autoref{tab:shapiro_test_all_features} are $<$ 0.05, and we reject the null hypothesis that these features are normally distributed. This is an important observation to guide our choices in selecting predictive models such as Random Forest and Decision Trees. Tree-based models perform well in utilizing non-normal features with inequalities in variance when predicting the target variable. This can be evidenced from our results when comparing models built with the feature set $\mathbf{X}$ both in \autoref{tab:results_author_centric}, and \autoref{tab:results_ext_agent}. Additionally, this suggests that parametric models like Multi-layer Perceptrons or Logistic Regression would only be advantageous if feature scaling is applied ($\mathbf{X_{scaled}}$) to normalize the features.

\begin{table}[H]
    \centering
    \small
    \begin{tabular}{lcc}
        \hline
        \textbf{Feature} & \textbf{Statistic} & \textbf{p-value} \\
        \hline
        Median Readability & 693.261011 & 2.885920e-151 \\
        Number of Algorithms & 43.248067 & 4.062576e-10 \\
        Number of Equations & 15.267781 & 4.837751e-04 \\
        Google Scholar citations & 35.751811 & 1.724221e-08 \\
        Measure of lexical textual diversity & 94.078257 & 3.725342e-21 \\
        \hline
    \end{tabular}
    \caption{Kruskal-Wallis test on the author-centric framework.}
    \label{tab:kruskal_wallis_test_auth_centric_framework}
\end{table}

The results from Levene's test for homogeneity of variances in the author-centric framework \autoref{tab:levenes_test_auth_framework}, and the external-agent framework \autoref{tab:levenes_test_ext_agent_framework} indicate that all features, except lexical diversity, show statistically significant differences in the non-homogenous nature of features across groups. The significant results from Levene’s test in both frameworks for several features (particularly readability, algorithms, and equations) suggest that these features differ not just in their average values but also in their variability among different categories of papers. This could have implications for how these features influence the artifact and reproducibility assessments in scholarly papers in our dataset.

\begin{table}[H]
    \centering
    \small
    \begin{tabular}{lcc}
        \hline
        \textbf{Feature} & \textbf{Statistic} & \textbf{p-value} \\
        \hline
        Median Readability & 697.771459 & 6.386612e-151 \\
        Number of Algorithms & 54.607980 & 8.324174e-12 \\
        Number of Equations & 28.063838 & 3.521685e-06 \\
        Google Scholar citations & 142.053160 & 1.363764e-30 \\
        Measure of lexical textual diversity & 108.002775 & 2.952022e-23 \\
        \hline
    \end{tabular}
    \caption{Kruskal-Wallis test on external-agent framework.}
    \label{tab:kruskal_wallis_test_ext_agent_framework}
\end{table}

We used significance tests such as the Kruskal-Wallis test to make statistical inferences about the variability of feature values across papers grouped by the author-centric and external-agent frameworks. Since our numerical features are not normally distributed, it is suitable to use a non-parametric test like Kruskal-Wallis. The results from \autoref{tab:kruskal_wallis_test_auth_centric_framework} and \autoref{tab:kruskal_wallis_test_ext_agent_framework} indicate significant differences for all the numerical features across groups of papers in both frameworks. The low $p$ values ($<$ 0.05) suggest that these features are valuable for predictive models, as their variability can help distinguish papers from different parts of the spectrum.
In summary, these results support our intuition that structural, linguistic, and scholarly features are useful for predicting artifact quality and reproducibility assessment status.

\section{Predictive Models}

Our goal is to build interpretable predictive models to estimate the reproducibility of scientific research. We develop two distinct multi-class predictive models, $\phi_{author}$ and $\phi_{external}$, to predict the label (e.g., $E_{R}$) of a paper in the author-centric and external-agent frameworks. We experimented with several predictive models. The results from the Shapiro test in \autoref{tab:shapiro_test_all_features} indicated that tree-based models such as Gradient Boosting, AdaBoost, Random Forest, and Decision Tree algorithms were more suitable. Non-parametric models such as Logistic Regression and Neural Networks were also used after applying a simple feature scaling technique using the mean and standard deviation.

The remarkable effectiveness of feature representations from large language model embeddings cannot be overstated. By using document representations from text-embedding models such as Davinci from OpenAI, and SPECTER and Longformer from AllenAI, we can capture the full semantic context of scholarly texts. Since scholarly documents often exceed the maximum sequence length allowed by these models, we split the documents and took the average of the embeddings as the final representation. We used two models for these representations: 1. A \textbf{VanillaNN}, which is a linear classifier, and 2. An \textbf{MLP} (multi-layer perceptron) with a hidden layer.

\subsection{Results for Author-Centric Framework}

We evaluate the effectiveness of our predictive models for the author-centric framework labels using classification metrics such as accuracy and F1 scores. The results are presented in \autoref{tab:results_author_centric}. 
 As mentioned in Section \ref{sec:spectrum}, the $\phi_{author}$ models predict one of three labels: $A_{PWA}$ (papers without artifacts), $A_{PUNX}$ (papers with artifacts that aren't permanently archived), and $A_{PAX}$ (papers with artifacts that are permanently archived). While it might seem that extracting artifact locations from paper texts would make a predictive model unnecessary, our experiments show that features designed to extract such information are not the best predictors. This highlights that predicting artifact availability or quality is a more challenging task than it appears.

The tree-based models, including Gradient Boosting, AdaBoost, Random Forest, and Decision Tree, demonstrate strong performance on the original feature set $\mathbf{X}$, with accuracy scores ranging from 78\% to 83\% and macro-averaged F1 scores between 66 \% and 74 \%. These results demonstrate the effectiveness of machine learning algorithms in distinguishing between papers with different levels of artifact availability, which is a critical aspect of reproducibility. In particular, the high F1 scores for $A_{PUNX}$ and $A_{PWA}$ indicate that these models are able to accurately differentiate between papers with and without permanently archived artifacts. On the other hand, non-parametric models like Logistic Regression and VanillaNN applied to the scaled feature set $\mathbf{X}_{\text{scaled}}$ show relatively weaker performance, which may be attributed to the loss of information during feature scaling. Finally, models leveraging text embeddings show promising results, particularly the MLP model with the ADA-002 embeddings, which achieves an accuracy score of 85\% and a macro-averaged F1 score of 77\%.

\begin{table*}[h]
    \centering
    \small
    \begin{tabular}{lcccccc}
    \hline
    \textbf{Model} & \textbf{$Acc$} & \textbf{$F_1(A_{PWA})$} & \textbf{$F_1(A_{PUNX})$} & \textbf{$F_1(A_{PAX})$} & \textbf{$F_1(macroavg)$} & \textbf{$F_1(weightedavg)$} \\ \hline
    $\mathbf{X}$ &  &  &  &  &  & \\
    Gradient Boosting & 0.83 & 0.82 & 0.89 & 0.52 & 0.74 & 0.82 \\
    AdaBoost & 0.78 & 0.77 & 0.86 & 0.34 & 0.66 & 0.76 \\
    Random Forest & 0.83 & 0.75 & 0.90 & \textbf{0.57} & 0.74 & 0.82 \\
    Decision Tree & 0.79 & 0.74 & 0.87 & 0.53 & 0.71 & 0.79 \\
    
    \hline
    $\mathbf{X}_{\text{scaled}}$ &  &  &  &  &  & \\
    Logistic Regression & 0.71 & 0.14 & 0.84 & 0.37 & 0.45 & 0.66 \\
    VanillaNN & 0.78 & 0.66 & 0.86 & 0.54 & 0.69 & 0.78 \\
    
    \hline
    $\mathbf{X}_{\text{emb}}$ &  &  &  &  &  & \\
    SimpleNN - $\mathbf{X}_{\text{emb(ADA-002)}}$ & 0.80 & 0.76 & 0.86 & 0.36 & 0.67 & 0.77 \\
    SimpleNN - $\mathbf{X}_{\text{emb(SPECTER)}}$ & 0.68 & 0.32 & 0.83 & 0.26 & 0.47 & 0.65 \\
    SimpleNN - $\mathbf{X}_{\text{emb(Longformer)}}$ & 0.83 & 0.97 & 0.89 & 0.08 & 0.65 & 0.67 \\
    MLP - $\mathbf{X}_{\text{emb(ADA-002)}}$ & 0.81 & 0.83 & 0.88 & 0.51 & 0.74 & 0.81 \\
    MLP - $\mathbf{X}_{\text{emb(SPECTER)}}$ & 0.68 & 0.29 & 0.82 & 0.33 & 0.48 & 0.66 \\
    MLP - $\mathbf{X}_{\text{emb(Longformer)}}$ & \textbf{0.85} & \textbf{0.97} & \textbf{0.90} & 0.43 & \textbf{0.77} & \textbf{0.83} \\
    \hline
    \end{tabular}
    \caption{Evaluation metrics for models predicting the author-centric framework labels.}
    \label{tab:results_author_centric}
\end{table*}

\subsection{Results for External-Agent Framework}

The results for models predicting the external agent framework labels are presented in \autoref{tab:results_ext_agent}. As mentioned in Section \ref{sec:spectrum}, the models here predict one of four labels: $E_{NR}$ (papers that cannot be reproduced), $E_{AR}$ (papers awaiting reproducibility), $E_{Re}$ (reproduced papers), and $E_{R}$ (reproducible papers). Overall, the best-performing model is an MLP that uses Longformer embeddings, which achieved the highest accuracy of 79\%, along with comparably high F1 overall scores, and individual class-specific scores. However, parametric models that used scaled features $\mathbf{X_{scaled}}$ demonstrated minimal predictive advantage of representational learning models. 

The tree-based models, such as Gradient Boosting, Random Forest, and Decision Tree, continued to perform well with accuracy scores ranging from 69\% to 75\% and macro-averaged F1 scores between 67\% and 72\%. Although these models are effective in distinguishing between papers that cannot be reproduced ($E_{NR}$) and papers awaiting reproducibility ($E_{AR}$), improvements are needed in predicting reproduced $E_{Re}$ and reproducible $E_{R}$ papers. The key takeaway from \autoref{tab:results_ext_agent} is the superior performance of models using embeddings ($\mathbf{X_{emb}}$), particularly those based on Longformer and ADA-002, compared to both basic models ($\mathbf{X}$) and those using scaled features ($\mathbf{X_{scaled}}$). Although this suggests that the semantic understanding provided by these embeddings is crucial in discerning subtle differences in paper statuses related to reproducibility, further investigation about reliability and robustness in predictions is necessary to fully understand model confidence (Section \ref{sec:label-conf}).

\begin{table*}[h]
    \centering
    \small
    \begin{tabular}{lccccccc}
    \hline
    \textbf{Model} & \textbf{$Acc$} & \textbf{$F_1(E_{NR})$} & \textbf{$F_1(E_{AR})$} & \textbf{$F_1(E_{Re})$} & \textbf{$F_1(E_{R})$} & \textbf{$F_1(macroavg)$} & \textbf{$F_1(weightedavg)$} \\ \hline
    $\mathbf{X}$ &  &  &  &  &  &  &  \\
    Gradient Boosting & 0.73 & 0.81 & 0.78 & 0.60 & 0.66 & 0.71 & 0.73 \\
    AdaBoost & 0.57 & 0.72 & 0.59 & 0.24 & 0.60 & 0.54 & 0.59 \\
    Random Forest & 0.75 & 0.74 & 0.81 & \textbf{0.63} & 0.68 & 0.72 & 0.75 \\
    Decision Tree & 0.69 & 0.77 & 0.76 & 0.57 & 0.58 & 0.67 & 0.69 \\
    
    \hline
    $\mathbf{X}_{\text{scaled}}$ &  &  &  &  &  &  &  \\
    Logistic Regression & 0.55 & 0.07 & 0.66 & 0.15 & 0.53 & 0.35 & 0.50 \\
    VanillaNN & 0.70 & 0.69 & 0.78 & 0.60 & 0.59 & 0.66 & 0.70 \\
    
    \hline
    $\mathbf{X}_{\text{emb}}$ &  &  &  &  &  &  &  \\
    SimpleNN - $\mathbf{X}_{\text{emb(ADA-002)}}$ & 0.75 & 0.79 & 0.81 & 0.44 & 0.68 & 0.68 & 0.74 \\
    SimpleNN - $\mathbf{X}_{\text{emb(SPECTER)}}$ & 0.57 & 0.30 & 0.70 & 0.38 & 0.54 & 0.48 & 0.57 \\
    SimpleNN - $\mathbf{X}_{\text{emb(Longformer)}}$ & 0.73 & 0.97 & 0.77 & 0.13 & 0.59 & 0.62 & 0.70 \\
    MLP - $\mathbf{X}_{\text{emb(ADA-002)}}$ & 0.74 & 0.83 & 0.81 & 0.52 & 0.63 & 0.70 & 0.74 \\
    MLP - $\mathbf{X}_{\text{emb(SPECTER)}}$ & 0.54 & 0.35 & 0.68 & 0.40 & 0.47 & 0.47 & 0.55 \\
    MLP - $\mathbf{X}_{\text{emb(Longformer)}}$ & \textbf{0.79} & \textbf{0.97} & \textbf{0.82} & 0.60 & \textbf{0.70} & \textbf{0.77} & \textbf{0.79} \\
    \hline
    \end{tabular}
    \caption{Evaluation metrics for models predicting the external agent labels.}
    \label{tab:results_ext_agent}
\end{table*}

\subsection{Important features for $\phi_{author}$ and $\phi_{external}$}

One of the contributions of this study is the identification of features that correlate well with the reproducibility of a paper. As shown in \autoref{tab:results_author_centric} and \autoref{tab:results_ext_agent}, the Random Forest model consistently performs best in terms of both accuracy and overall F1 score across both frameworks. As a result, we selected this model for further analysis in the feature importance study.
\begin{figure*}[htbp]
    \centering
    \includegraphics[width=0.9\textwidth]{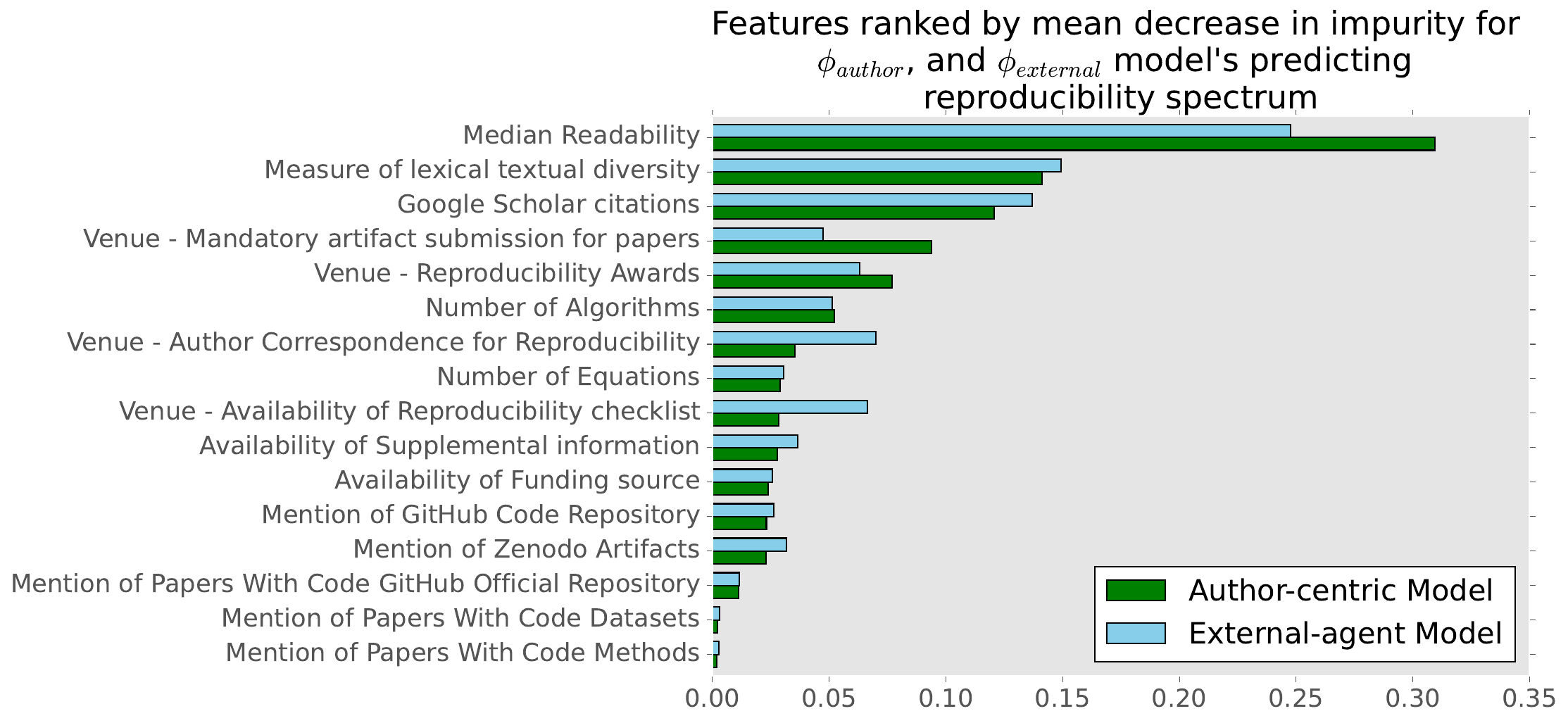}
    \caption{Most important features for predicting labels in the author-centric, and external-agent frameworks.}
    \label{fig:imp_features_joint}
\end{figure*}
We collected the Gini impurity importance for all features in the Random Forest model (in both frameworks) and ranked them in Fig.~\ref{fig:imp_features_joint}. 
Linguistic measures such as readability and lexical diversity strongly influence the predictive outcomes of the models. Intuitively, clarity in language and thoroughness in explaining concepts (modeled through readability and lexical diversity features) should neither be correlated with the quality of artifacts nor should it affect the reproducibility status of a paper. However, the influence of these features on the predictive models, especially Random Forest, suggests otherwise. This surprising finding has also been observed in previous studies \cite{guerini2012linguistic, maclaughlin2018predicting, jin2021research}.

Among the top five features, we also observe the importance of citations and other venue-based features in both models. Citations act as a latent variable connecting a scholarly paper's impact and credibility. Highly cited papers might be considered more reproducible due to peer validation, but results from \cite{yang2020estimating} suggest there is more room for introspection. The justification for having venue-based features, such as Reproducibility Awards, is to assess if such a variable serves the purpose of motivating authors to put more effort into making the artifacts available and consequently voluntarily opting in for reproducibility evaluation. Other categorical features that measure connections to references of supplemental information either within a paper or external sites such as Zenodo, Github, and PapersWithCode appear to have relatively lower rankings. Direct references to repositories where code and artifacts are stored are expected to be significant, given their role in facilitating artifact evaluation and reproducibility. However, the lower Gini importance suggests that additional factors are influencing the outcomes. Further research and experimentation are needed to uncover more latent variables within both frameworks.

\subsection{Model confidence for $\phi_{author}$ and $\phi_{external}$}
\label{sec:label-conf}

\begin{figure*}[htbp]
    \centering
    \includegraphics[width=0.49\textwidth]{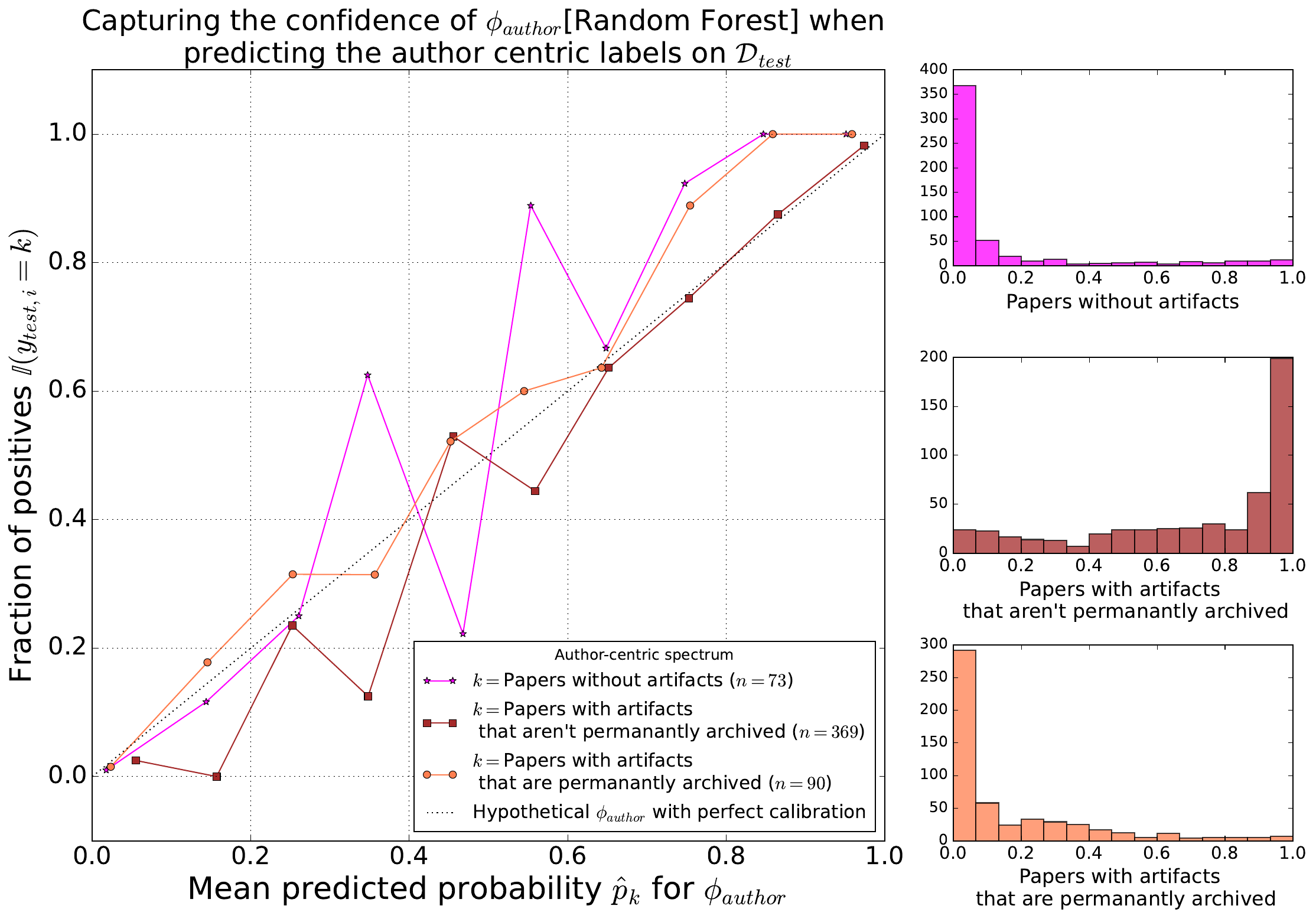}
    \hfill
    \includegraphics[width=0.49\textwidth]{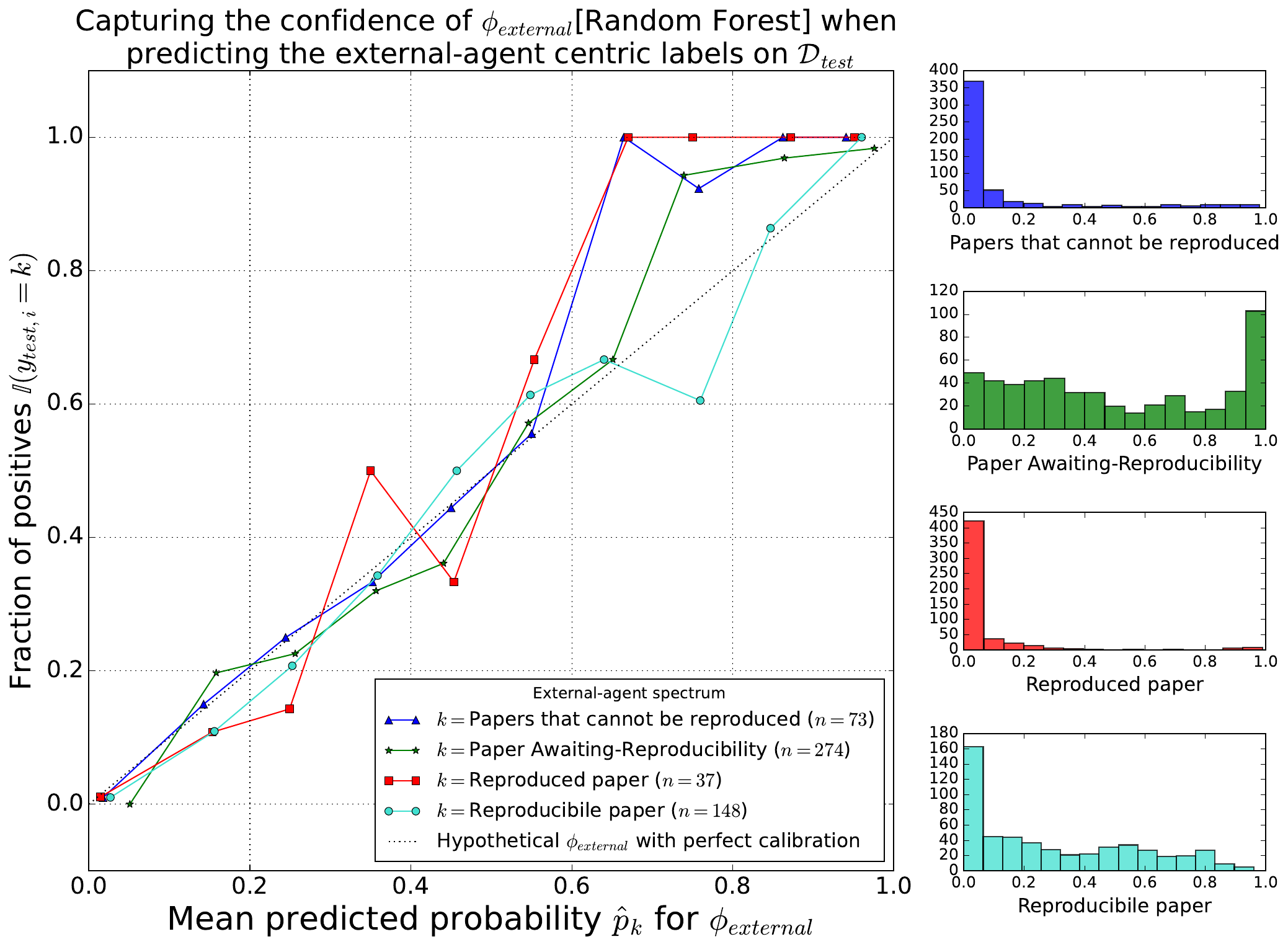}
    \caption{Confidence calibration of $\phi_{\text{author}}$ Random Forest model, author-centric framework (left) and $\phi_{\text{external}}$ Random Forest model, external-agent centric framework (right).}
    \label{fig:rf_author_centric_external_agent_model_conf}
\end{figure*}

\begin{figure*}[htbp]
    \centering
    \includegraphics[width=0.49\textwidth]{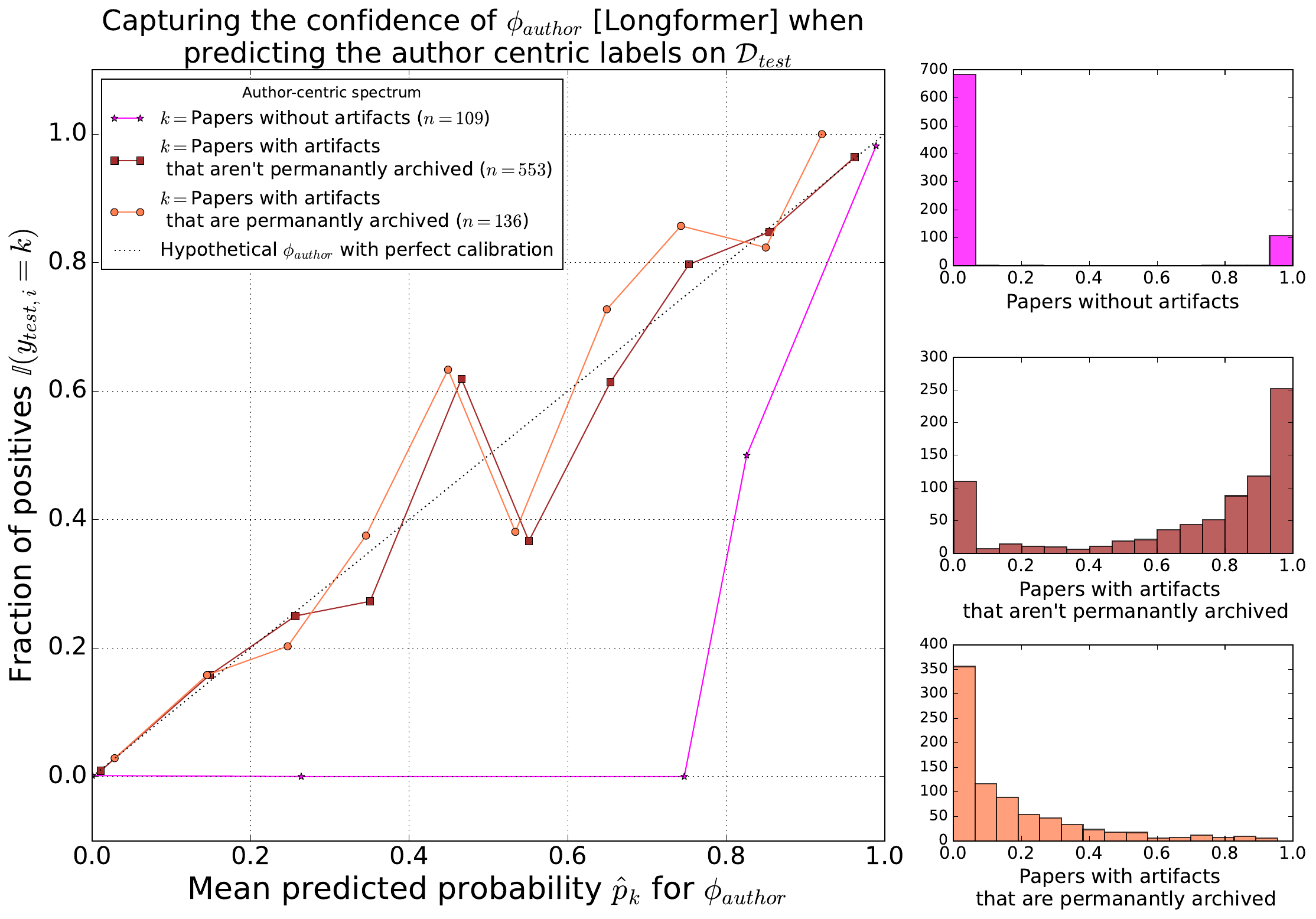}
    \hfill
    \includegraphics[width=0.49\textwidth]{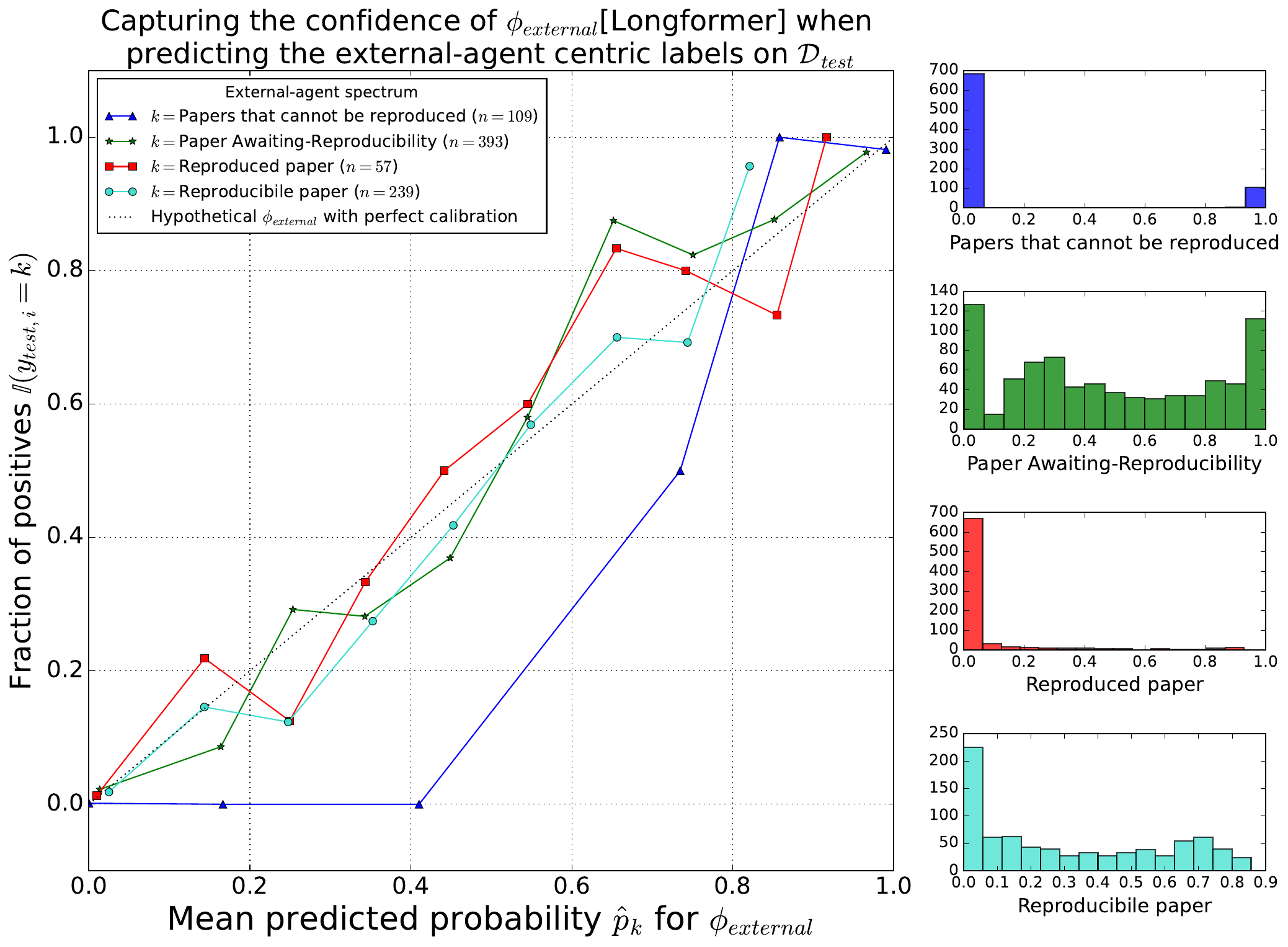}
    \caption{Confidence calibration of $\phi_{\text{author}}$ Longformer-MLP model, author-centric framework (left) and $\phi_{\text{external}}$ Longformer-MLP model, external-agent centric framework (right).}
    \label{fig:longformer_author_centric_external_agent_model_conf}
\end{figure*}

Understanding the confidence of predictive models is critical for establishing reliability. The confidence calibration curves for our models in are shown in Figs.~\ref{fig:rf_author_centric_external_agent_model_conf} and~\ref{fig:longformer_author_centric_external_agent_model_conf}. The bigger plots on the left show model confidence curves with mean predicted probabilities $\hat{p}_k = \frac{1}{n_k} \sum_{i=1}^{n_k} \hat{p}_{ik}$ on the $x$ axis, and fraction of positives observed through an indicator function $\mathbb{I}(y_{test, i} = k)$ on the y-axis, which evaluates whether the predicted category $k$ aligns with the actual category of each paper. In other words, these plots visualize the fraction of papers correctly identified within each category as a function of the predicted probabilities, allowing us to assess the calibration of the models across different categories of papers. 
The smaller plots on the right side of the confidence curves are histograms that show the overall distribution of predictive probabilities for each category of papers. These plots are useful for understanding the distribution of confidence the models have in their predictions. 
Fig.~\ref{fig:rf_author_centric_external_agent_model_conf} suggests that in the author-centric framework, a Random Forest model is reliable \textit{only} when it predicts if papers have permanently archived artifacts. Also, the mean predicted probabilities in the range 0.2--0.6 suggest it is not confident in predicting $A_{PWA}$, or $A_{PUNX}$. The Longformer model (\textit{MLP with longformer}, Fig.~\ref{fig:longformer_author_centric_external_agent_model_conf}) shows a weakness in reliability compared to the Random Forest model. It shows consistent under- or overconfidence across the author-centric labels, especially at higher probabilities for $A_{PWA}$. Most importantly, Fig.~\ref{fig:longformer_author_centric_external_agent_model_conf} suggests that despite its effectiveness in evaluation metrics, the Longformer model is less effective at assessing papers without artifacts, potentially due to a lack of distinguishing features in the embeddings. 

In the external-agent framework, for papers that cannot be reproduced ($E_{NR}$), we notice that Longformer model (Fig.~\ref{fig:longformer_author_centric_external_agent_model_conf}) is extremely under confident, predicting lower probabilities than the actual outcomes. Additionally, the confidence of the Longformer, when predicting papers awaiting reproducibility ($E_{AR}$), or Reproduced ($E_{Re}$), or Reproducible ($E_{R}$) papers, is variable, especially at higher probabilities, suggesting slight inconsistencies in predictive robustness, and reliability. The Random Forest model, on the other hand (Fig.~\ref{fig:rf_author_centric_external_agent_model_conf}) shows a better alignment in predictive probabilities against the fraction of positives for $E_{NR}$, $E_{AR}$, $E_{Re}$, and $E_{R}$. This suggests the Random Forest model is better when compared to an MLP with Longformer representations, specifically when we talk about reliability, robustness, and consistency of the labels predicted across both frameworks.
The histograms corroborate the reliability curves, indicating that the Random forest model not only predicts with high confidence but also aligns these predictions closely with the actual outcomes, which is critical for downstream applications using predictive models for analyzing reproducibility.

\section{Conclusion \& Future Work}
We define a spectrum to assess the reproducibility of scientific papers, collect a new dataset, and establish a framework for automatic prediction of the reproducibility of scientific papers. Our work presents a thorough analysis of predictive models that include feature importance tests and confidence calibration curves. We draw two surprising conclusions: 1. Linguistic features such as readability and lexical diversity are strong predictors for both the quality of artifacts mentioned in a paper and their reproducibility status, and 2. Neural nets built on text embeddings from large language models are accurate but not robust. 

This work can be improved and extended in various ways. The predictive models can be improved, and the Neural nets can be made more robust. The unreasonable effectiveness of linguistic features can be investigated. Using a model or algorithmically-driven intelligent system to reward ``reproducible'' research practices, however, can be problematic, and we must have foresight in developing an approach toward quantifying reproducibility to avoid potential ethical problems. For example, suppose a model or system finds that the language of a paper positively affects its likelihood to be reproducible. It may thus penalize research simply because of the language in which the paper is written. Similarly, a model or system could identify institutions it associates with more reproducible results. Then, papers submitted from that institution might be labeled by the model as reproducible, without considering their content. Certainly, these are not outcomes we would expect or desire of such an algorithm or model.
Code and data artifacts are critical for reproducibility evaluation, and papers without artifacts and papers that cannot be reproduced represent a sizeable portion of scientific literature. While it can be argued that features such as the Number of Algorithms, Equations, and Reproducibility checklists are aligned more toward ACM's Badging policy, the foundational principles of reproducibility are universal and not exclusive to ACM. The structure of computational science adopted by most researchers involves artifacts. These artifacts, when made available, enable other researchers to verify, build upon, and extend the original work. This process of verification and extension, facilitated by accessible artifacts, creates a pathway for more generalizable findings. Utilizing our spectrum through the author-centric and external-agent framework for a larger multi-disciplinary study will offer valuable insights into the broader landscape of scientific research reproducibility.
\textbf{Limitations:}
Generalizing the findings of our study to other disciplines is both data-intensive and challenging. While it is true that the composition of the ACM dataset and predictive modeling experiments cater to a specific category of computational science papers, the heuristics used to create the joint spectrum for reproducibility and the catalog of experiments we presented show a tangible pathway for expanding the study across other scientific disciplines. Despite the limitations, our work offers robust findings across the experiments, affirming the importance of ``readability" for reproducibility. 

\section*{Acknowledgement}
This work is supported in part by NSF Grant No. 2022443. The experiments involved in the study were run on Google Cloud, and compute is supported by the Google Cloud Research Credits program with the award 274000118.

\bibliographystyle{ACM-Reference-Format}
\balance
\bibliography{cikm_repro_paper}
\end{document}